\documentclass[fleqn,twoside]{article}
\usepackage{espcrc2}

\usepackage{graphicx}
\usepackage[figuresright]{rotating}

\title{Probing the Dark Ages}

\title{Strong Gravitational Lensing with SKA}
\author{L.V.E. Koopmans\address[koopmans]{Kapteyn Institute, P.O.Box
800, 9700AV Groningen, The Netherlands}, I.W.A. Browne$^{\rm b}$ and
N.J. Jackson\address[browne]{Jodrell Bank Observatory, Macclesfield,
Cheshire SK11 9DL, United Kingdom} }
   
\begin{document}

\begin{abstract}
The advent of new observational facilities in the last two decades has
allowed the rapid discovery and high-resolution optical imaging of
many strong lens systems from galaxy to cluster scales, as well as
their spectroscopic follow-up. Radio telescopes have played the
dominant role in the systematic detection of dozens of new
arcsec-scale lens systems. For the future, we expect nothing less!
The next major ground- and space-based facilities, especially the {\sl
Square Kilometer Array} can discover tens of thousands of new lens
systems in large sky surveys. For optical imaging and spectroscopic
follow-up a strong synergy with planned optical facilities is
needed. Here, we discuss the field where strong gravitational lensing
is expected to play the dominant role and where SKA can have a major
impact: The study of the internal mass structure and evolution of
galaxies and clusters to $z\sim 1$. In addition, studies of more
exotic phenomena are contemplated. For example, milli- and
microlensing can provide a way to measure the mass-functions of stars
and CDM substructure at cosmological distances. All-sky radio
monitoring will also rapidly develop the field of time-domain lensing.
\end{abstract}

%\begin{keyword}
%Gravitational Lensing
%\end{keyword}

\maketitle

\section{Introduction}

Gravitational lensing \cite{1992grle.book.....S} is the study of the
deflection of light-rays from their straight Euclidean path due to the
curvature of space-time caused by the presence of matter or energy.
Whereas only a curiosity several decades ago, gravitational lensing
has rapidly developed into a tool widely used in astrophysics and
cosmology.  It has already become the preferred method to study dark
matter in the Universe, it can be used to study mass scales from
planets ($\sim 10^{-4}$~M$_\odot$) to large-scale structure ($\gg
10^{14}$~M$_\odot$) and the conclusions do not depend on the nature of
the gravitating mass and its dynamical state. In addition, the large
redshift range over which lensing is seen allows one to measure the
evolution of properties of the dark and luminous mass distribution,
something that is difficult to do in any other way. Here, we discuss
the impact that SKA\cite{workshop} can have on {\sl strong
gravitational lensing} -- where multiple images of a single background
source are formed -- and in particular how these lens systems can be
used to study the structure and evolution of galaxies and clusters.
Some of this will be based on current understanding, but some will
necessarily be more speculative. We first set the scene by estimating
how many strong-lens systems SKA can discover, based on a large sky
survey. We limit the discussion of the more technical details -- which
have been discussed elsewhere \cite{Jackson_SKA} -- and focus on the
science with an ensemble of lensed systems three orders of magnitude
beyond what we have to date.

\section{The Radio All-SKA Lens Survey}

Based on the current specifications of SKA (October 23, 2003), a Radio
All-SKA Lens (RASKAL) Survey of half of the sky ($2\times 10^4$
sq. degrees) at 1.4~GHz can be accomplished in about five months
(10\,min per 1 sq. degree pointing, assuming a single beam.) to a
depth of several tenths of $\mu$Jy per beam with a resolution of
0.01$''$--0.02$''$. This survey will undoubtedly be done for many
reasons other than lensing. If we limit sources to brighter than
3~$\mu$Jy\footnote{A signal-to-noise $\ge$10 per beam will be
sufficient for a clear identification of most sources when being
lensed. However, since most sources are close to the survey flux limit
and every pointing encompasses many lensed systems, deeper follow-up
of {\sl all} faint lensed systems will be equivalent to an all-sky
survey deeper than 10 min per pointing. For a $>$1~sq.~degree FOV or
multiple beams, the effective integration per pointing could be
$>$10~min, increasing the number of lensed systems.}, this survey
yields $\sim 10^5$ sources per square degree or $10^9$ sources per
hemisphere. Based on the lensing optical depth found from the CLASS
radio lens survey \cite{CLASS1,CLASS2}, $\tau=(1.4\pm0.4)\cdot
10^{-3}$, one might expect the number of lensed sources to be $N_{\rm
lens}\sim 10^6$ if the redshift distributions of the CLASS and SKA
sources are similar. Most of these sources are expected to be extended
star-burst galaxies, but $\sim$10\% might still be compact
flat-spectrum AGNs. The latter, as was the case for CLASS, are easy to
identify and yield a number of lensed systems $\sim 10^5$, assuming
these AGN are mostly at $z>1$. The extended sources will appear as
arcs and rings, with curvature radii ranging between a few tenths and
several arcsec for galaxies and potentially tens of arcsec for
clusters. The dominant uncertainty is the typical redshift of
starbursters. For sources $>10 \mu$Jy, the median redshift is expected
to be $z>1$ and $\tau\sim 10^{-3}$ for $\sim 10^8$ available sources
\cite{Jackson_SKA}.

\begin{figure}[ht]
\includegraphics[width=3in]{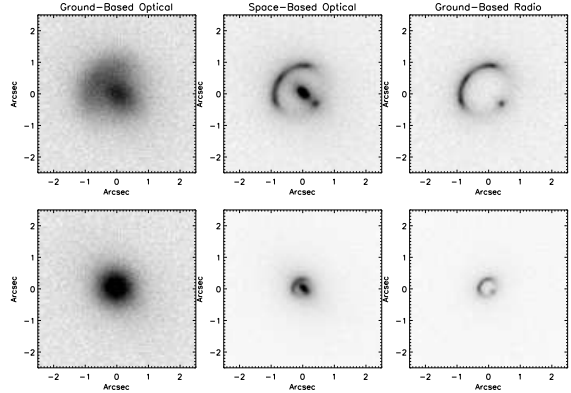}
\vspace{-1cm}
\caption{A lens system with a source lensed into a four-image
arc/ring. The upper (lower) panels show an early-type lens with an
Einstein radius of $R_{\rm E}=1.2(0.4)''$. From left, middle to right,
the FWHM resolution decreases from 0.7$''$, 0.1$''$ to 0.02$''$ (e.g.\
expected for LSST, SNAP, SKA, respectively). The two right panels show
the radio system, as observed with SKA, without the lens
galaxies. Note that in the optical the lens-galaxy significantly
contaminates the emission from the lensed source, making it harder to
identify these systems as lenses, whereas in the radio even
small-separation systems are relatively easier to identify. [The lens
and source have $R^{1/4}$ brightness profiles with effective radii of
1.5$''$.  The lens is 5 times brighter than the un-lensed source. The
S/N=1 per pixel (0.05$''$) at the effective radius of the lens for
each panel (the brightness range is set differently to bring out the
structure).]\label{fig:sims}}
\end{figure}

In the following we therefore assume $\sim 10^5$ lensed systems with compact
flat-spectrum AGN sources and at least $>10^5$ with extended starburst
sources. We should note that the actual {\sl parent population} of the
lensed sources can be $\sim$10 times fainter, and thus more numerous,
which increases the above lens-rates. However, the SDSS has taught us
that estimating the number of expected lensed systems is easier than
identifying them in the actual data.  In the following section, we
discuss why is SKA is expected to do better in identifying new
lens systems.

\section{Identification and Expectation of New Lensed Systems}

To identify $\sim 10^{5}$ lensed systems, special search strategies
have to be developed. Whereas for $\sim 10^4$ sources with two dozen
lensed systems (e.g.\ CLASS), a search-by-eye is feasible, it clearly
is not for $\sim 10^9$ sources. There are several advantages of
searching for new lensed systems in the radio, that make this task
easier than e.g. in optical surveys: (i) To clearly identify a lens
system, at least several resolution elements across the system are
needed. The typical image splitting is $\sim$1.2$''$ for L$_*$ galaxy
lenses with $\sigma\approx 225$\,km\,s$^{-1}$, hindering ground-based
optical surveys with comparable seeing (e.g.\ SDSS). On the other
hand, SKA has an anticipated resolution of $0.01''-0.02''$ at
1.4\,GHz, more than sufficient to identify even small dwarf-galaxy
lenses with $\sigma\approx 50$\,km\,s$^{-1}$ (e.g.\ Figure 1.)  (ii)
At 1.4 GHz the emission of a non-starburst early-type lens galaxy
(i.e.\ the dominant lens) is often small compared to the source
emission\footnote{Note that some lens galaxies have radio-bright
AGN. The latter, however, are typically compact and seen in the galaxy
cores, where emission of the lensed source is strongly demagnified.}.
This improves chances of identifying a source as being lensed, in
particular if the source is poorly resolved. (iii) The bandwidth of
$\Delta\nu/\nu\approx0.25$ allows disentanglement of complex multiple
images (e.g.\ rings and arcs) based on their spatial spectral-index
distribution. (iv) Unlike optical surveys, in the radio lensed images
are not affected by dust-extinction, although the most compact $\mu$Jy
sources might scintillate. The latter, however, remains limited at 1.4
GHz due to the small (10\,min) integration times and its effect can be
assessed from its frequency-dependent behavior within the wide
bandwidth.

In the case of compact flat-spectrum AGNs, one can pre-select
candidates using only the longest baselines, such that extended
emission is resolved out. Multiple compact images within several
arcsecond (i.e. galaxy scale) are a good indication of a lensed
system, in particular if their low-frequency radio spectra are similar
(at $\nu\sim1$~GHz, these sources are not expected to vary
strongly). The case of extended sources is more complex and requires
more sophisticated analyzes of the data, for example through neural
networks. The high resolution, relatively minor contamination by
emission from the lens galaxy and large frequency coverage all
contribute to making this task easier, although it remains difficult
to assess the completeness of such a strategy.
We can neither expect to find {\sl all} lensed systems, nor expect an ensemble
to contain only {\sl genuine} lensed systems. Any algorithm that is more
restrictive reduces the number of non-lenses in the ensemble, but also
also rejects genuine lensed systems\footnote{Note that a by-eye check of
$10^5$ lensed systems would take $\sim$4 eight-hour work days, spending only
1~second per lens!}. However, {\sl ``completeness regions''} might be
identified for statistical purposes (e.g.\ studying galaxy evolution
and cosmography).

Assuming that we find $>10^{5}$ lensed systems, what type of
strong-lens systems can we expect? Simply speaking, all objects with a
density greater than $\Sigma_{\rm crit}=3.5 \hat{D}^{-1}$~kg~m$^{-2}$
with $\hat{D}=D_{\rm d} D_{\rm ds}/D_{\rm s}$ (distances are in Gpc)
can in principle multiply image a source. This includes nearly all
collapsed objects in the Universe: {\sl asteroids, comets, planets,
brown dwarfs, stars, stellar remnants, primordial black holes,
globular clusters, galaxies, some compact groups, and clusters}.
Whereas at distances of several kpc only the most dense objects can
multiply image (e.g.\ asteroids, \dots, primordial black holes) at
cosmological distances of Gpc, the less dense objects (e.g.\ galaxies,
\dots, clusters) become the dominant observable lenses. This is
because the typical image separation is $\theta_{\rm Einst}=3
(M/M_\odot) \tilde{D}^{-1/2}$~$\mu$as, with $\tilde{D}=D_{\rm d}
D_{\rm s}/D_{\rm ds}$, which for stars is of order micro-arcsec and
for galaxies and clusters arcsec to arcmin. Whereas current telescopes
are limited by their resolution in their ability to find lenses much
smaller than galaxies at cosmological distances, SKA with 0.02$''$
resolution can discover lensing by objects as small as dwarf galaxies
($>10^{7}$~M$_\odot$) or even smaller if SKA is combined with other
telescopes in VLBI mode. Even so, the dominant type of lens is
determined by the total mass fraction in those objects. Since the
total number of lenses is $N_l\propto \Omega_l/M_l$ and the lens
cross-section $\sigma_l \propto \theta_{\rm Einst}^2$, the total
integrated cross-section of the lens population is
$\sigma_{l,tot}\propto \Omega_l$. Hence, structures like galaxies and
clusters that dominate $\Omega_{\rm m}$, will also dominate the
population of lens systems. We therefore think that their study will
be one of the largest beneficiaries of an all-sky strong lens survey.

\section{The Internal Structure \& Evolution of Galaxies and Clusters}

Galaxies and clusters are the most massive collapsed, possibly
relaxed, structures in the Universe. Their internal structure and
evolution provide clues to how the matter distribution evolves from
the linear to highly non-linear regime. Until recently, the mass
distribution and evolution of galaxies and clusters have only be
studied through their luminous mass, e.g. HI rotation curves, polar
rings, X-ray observations and stellar dynamics. Degeneracies in these
techniques and the required high--S/N data have limited such studies
to the local Universe (i.e.\ $z<0.1$).

In the more distant Universe, gravitational lensing provides a
practical tool to study their inner mass structure in more
detail. Many of the lens systems -- to be discovered by SKA -- show
extended arcs and rings (Figure 2), providing more powerful
constraints on the lens potential.  Combined constraints with other
techniques (e.g.\ stellar dynamics and high-resolution optical
imaging) can help to further disentangle the distributions of luminous
stellar mass from the dark-matter halo in early-type galaxies out and
beyond $z=1$. Similar techniques can be used for spiral galaxies and
clusters. We now discuss these separately in somewhat more detail.

\begin{figure}[t]
\begin{center}
\hbox{
\resizebox{0.44\hsize}{!}{\includegraphics{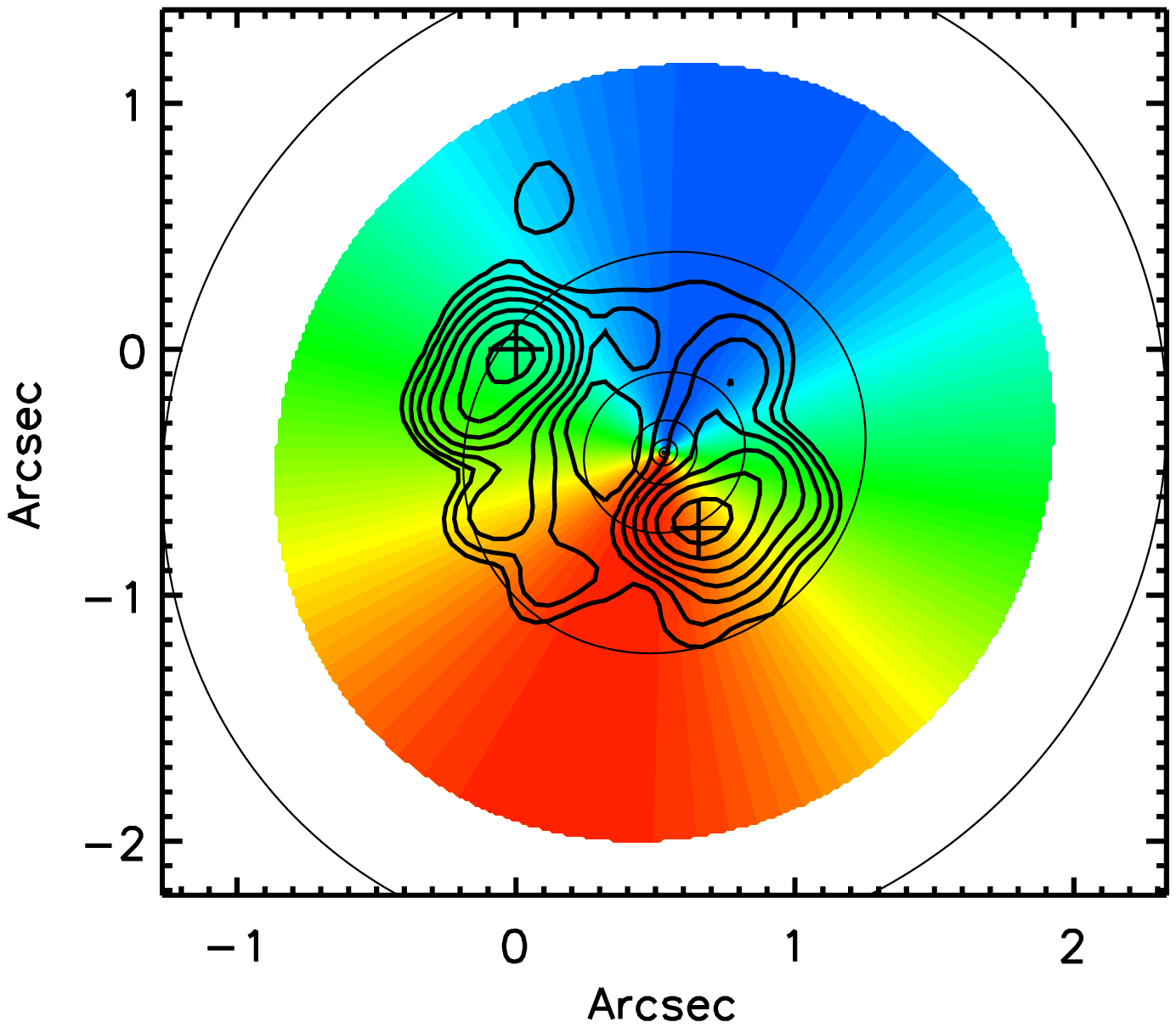}}
\resizebox{0.55\hsize}{!}{\includegraphics{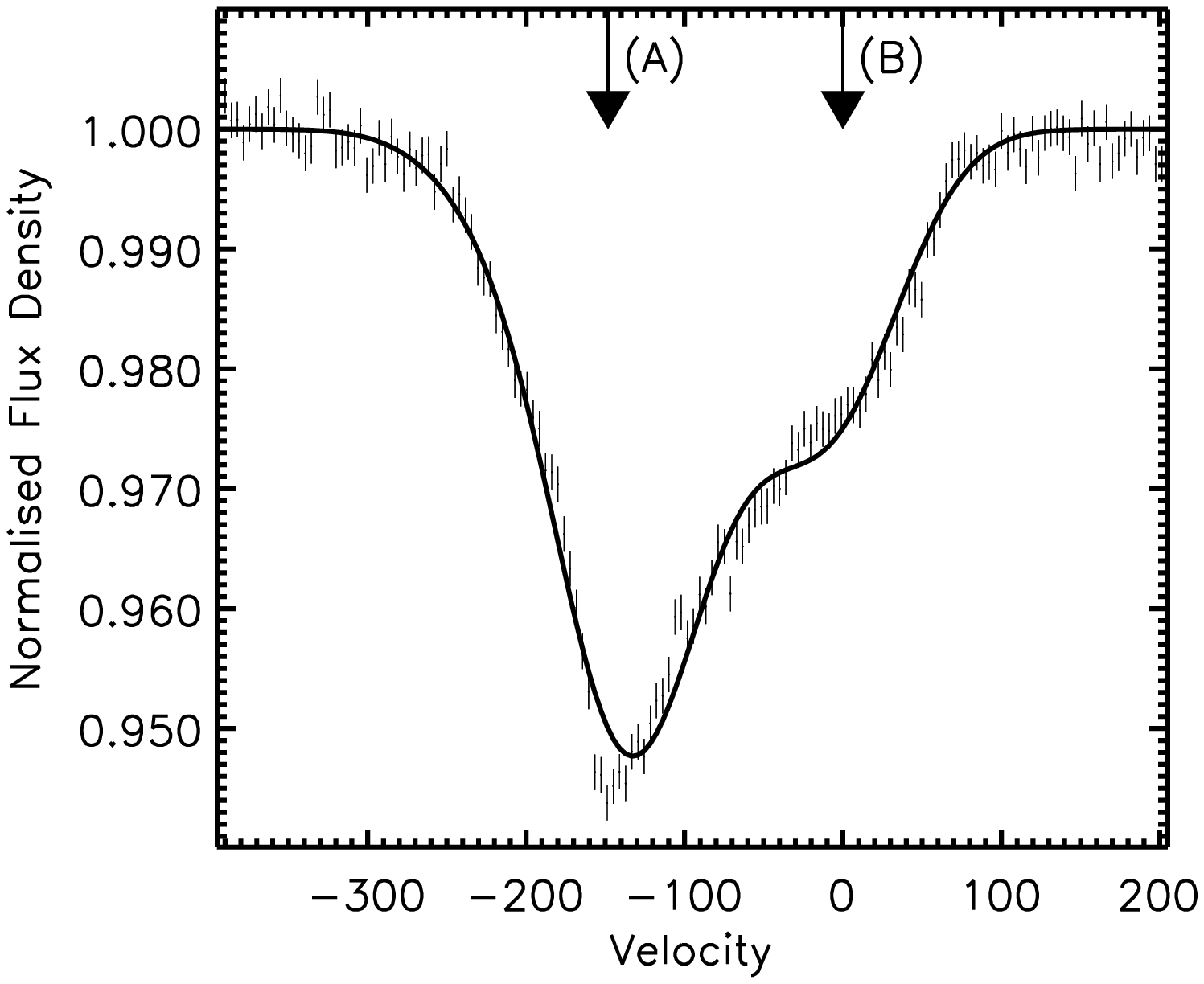}}}
\vspace{-0.5cm}
\parbox[b]{\hsize}{
\caption{{\bf Left:} A MERLIN 1.4--GHz radio image (contours) of
PKS1830--211 overlaid on a model of the HI velocity field in the
spiral lens galaxy at z=0.89. {\bf Right:} An integrated WSRT HI
absorption line, overlaid with the best-fit model of the line, given
the HI velocity field shown in the left panel.
\label{fig:1830}}}
\end{center}
\end{figure}

\subsection{Disk-dominated Galaxies}

The determination of the mass distribution of disk galaxies has
historically been dominated by the study of their rotation curves,
sometimes supported by stellar-kinematic data. Even though rotation
curves are easy to translate into enclosed mass, it remains nearly
impossible to determine their stellar $M/L$ ratios with confidence. In
particular at high redshifts, little or no data is available. SKA will
offer several opportunities to measure mass profile of disk-dominated
galaxies. First, through rotation curves of the more HI-rich disk
galaxies to $z\sim$1. However, several equally exciting opportunities
are offered by gravitational lensing\footnote{Although lens-galaxies
are dominated by early-type galaxies, we expect to find at least $\sim
10^{4}$ disk-galaxy lenses with $L>0.01\,L_*$ (assuming an image
separation $\Delta\theta=0.04''$ is sufficient to recognize a lens and
$\Delta\theta_*=0.4''$, with $L\propto
\sigma^4\propto\Delta\theta^2$).}: (i) Direct measurement of the total
mass profile through lensing-constraints from extended arcs and
complete Einstein rings (e.g.\ B0218+357). (ii) HI-absorption
measurements against a lensed radio source, combined with the lensing
constraints from the rings/arcs themselves (e.g. PKS1830-211; Figure
2).  This provides complementary lensing and kinematic
information. Since the brightness of the source does not correlate
with the HI content of the lens galaxy, this technique can be used
even for dwarf galaxies that are too faint in HI emission.  For
hundreds of disk galaxies with HI gas and relatively bright background
sources, deeper follow-up can be done, to map their kinematic field
over the extent of the lensed arcs/rings. A study of disk galaxies to
$z>1$ and possibly as faint as $0.01\,L_*$ (i.e.\ dispersions
$\sigma=v/\sqrt{2}>50\,$km/s) can thus be imagined.

\subsection{Early-type Galaxies}

For early-type lens galaxies very similar studies can be done.
However, the complementary kinematic data is obtained from optical
spectra obtained with large-aperture telescopes, not from HI
gas. Recent studies of early-type galaxies at $z=0.5-1.0$ have shown
that gravitational lensing -- especially the enclosed mass of the
galaxy -- allows the mass-anisotropy degeneracy in stellar dynamics to
be broken. Constraints can then be set on the mass fraction and inner
slope of their dark-matter halos, providing direct evidence for dark
matter around these galaxies beyond the local Universe.

The gain is obvious with $\sim10^5$ new radio-selected lens systems to
choose from, instead of the current $\sim$100 systems of which only a
fraction is useful for detailed optical and kinematic
follow-up\footnote{Lens surveys that target bright optical sources
(e.g.\ quasars) often limit detailed studies of the fainter lens
galaxies.}. Properties of the luminous and dark-matter distributions
can be studied over a much wider range of parameter space and also in
time (i.e.\ redshift), allowing evolution of the mass distribution of
galaxies to be studied directly. Whereas most cosmological studies
thus far focus on the evolution of galaxies in terms of their baryonic
content (e.g.\ gas and stars), little to nothing is known about their
mass and structural evolution.  Strong gravitational lensing with the
large samples that SKA can discover will change this situation.

\subsection{Clusters of Galaxies}

Strong lensing by clusters, although more massive, will be less
common. However, with a lensing rate of $\sim$1:50,000 we still expect
thousands of strong lens cases by clusters to be discovered in a
RASKAL Survey. Since this is a ``blind'' survey (i.e.\ one targets the
sources, not the lenses), a sample of cluster lenses can be used to
accurately quantify their mass-function with redshift. Such a sample
would also be an indicator of the evolution of structure from the
linear to non-linear scales, a strong function of the cosmological
model. Deep integrations on individual clusters (e.g.\ identified from
deep optical surveys) can unveil a wealth of radio-arcs from the
population of extended star-burst galaxies at the $\mu$Jy level, used
to map the inner mass distribution of these clusters in detail. Also
weak-lensing studies, beyond the scope of this paper, can be done at
par with optical studies (see contribution by Blake et al. in this
volume). The great advantage of deep cluster surveys with SKA is that
foreground (including cluster) galaxies do not severely ``hinder'' a
study of the lensed background, since they are often relative faint in
the radio (i.e.\ mostly early-type galaxies).

\begin{figure*}[t!]
%\begin{center}
\vbox{
\hbox{ 
\resizebox{0.3\hsize}{!}{\includegraphics{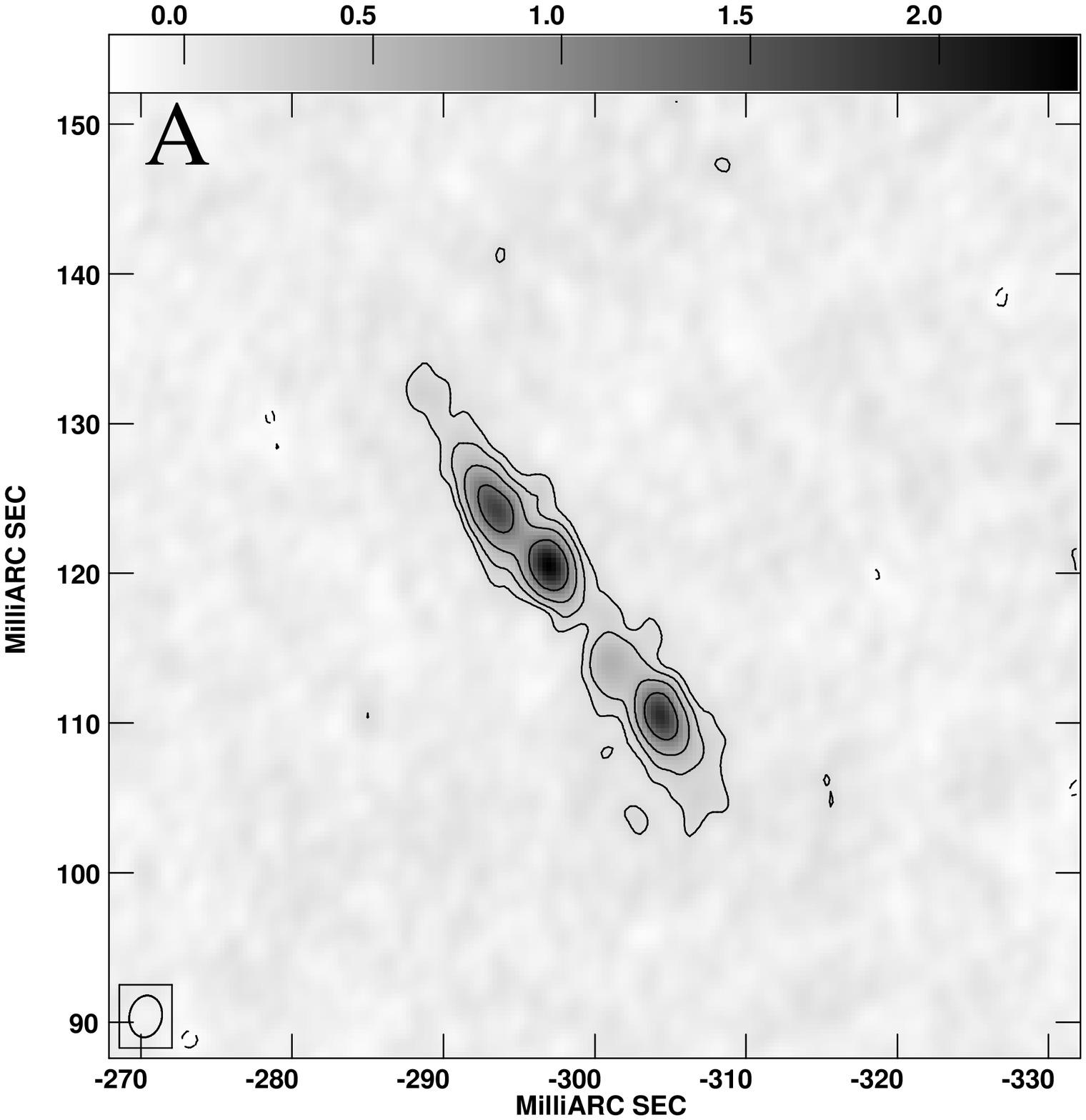}}
\resizebox{0.3\hsize}{!}{\includegraphics{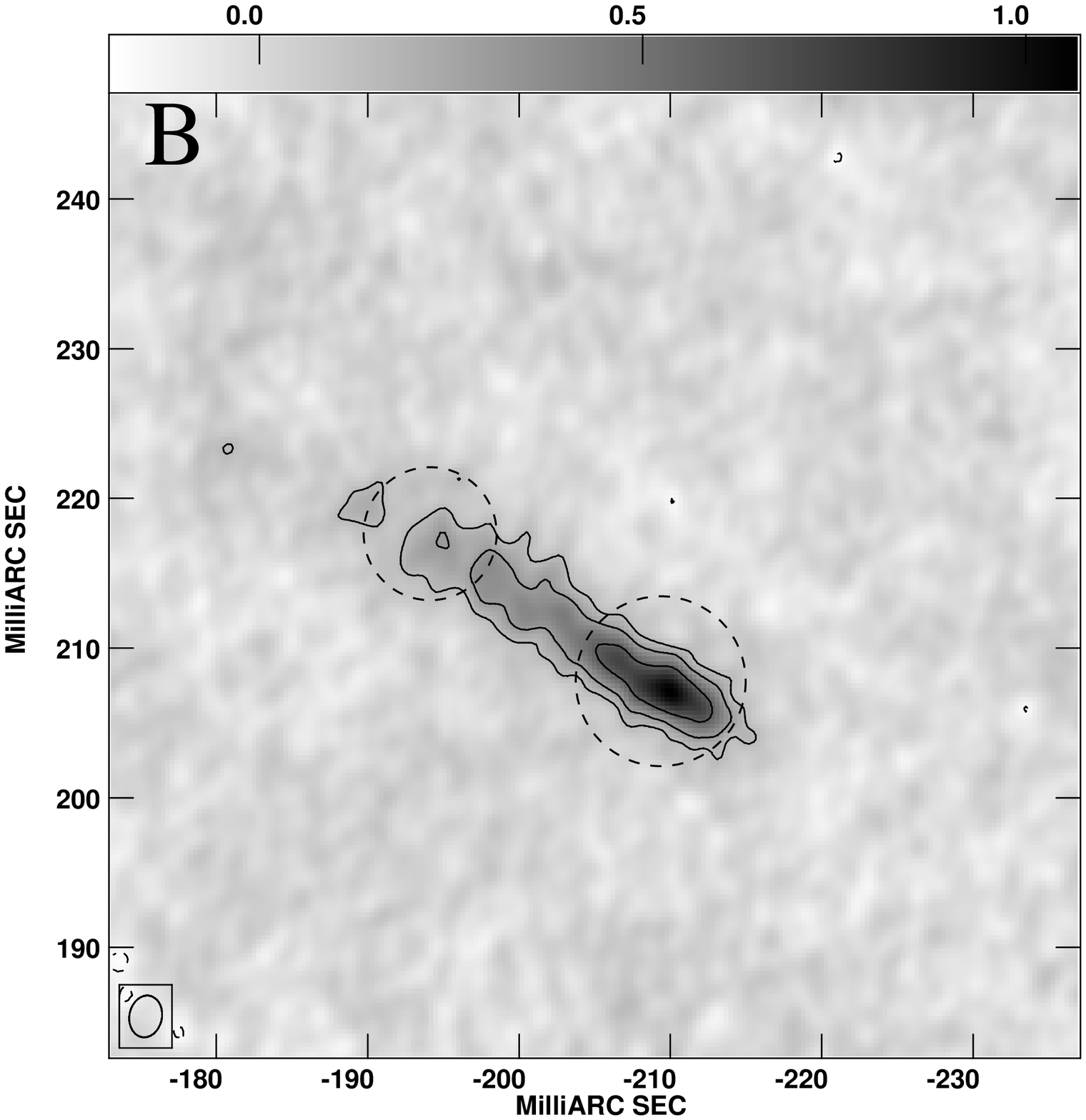}}}
\hbox{ 
\resizebox{0.3\hsize}{!}{\includegraphics{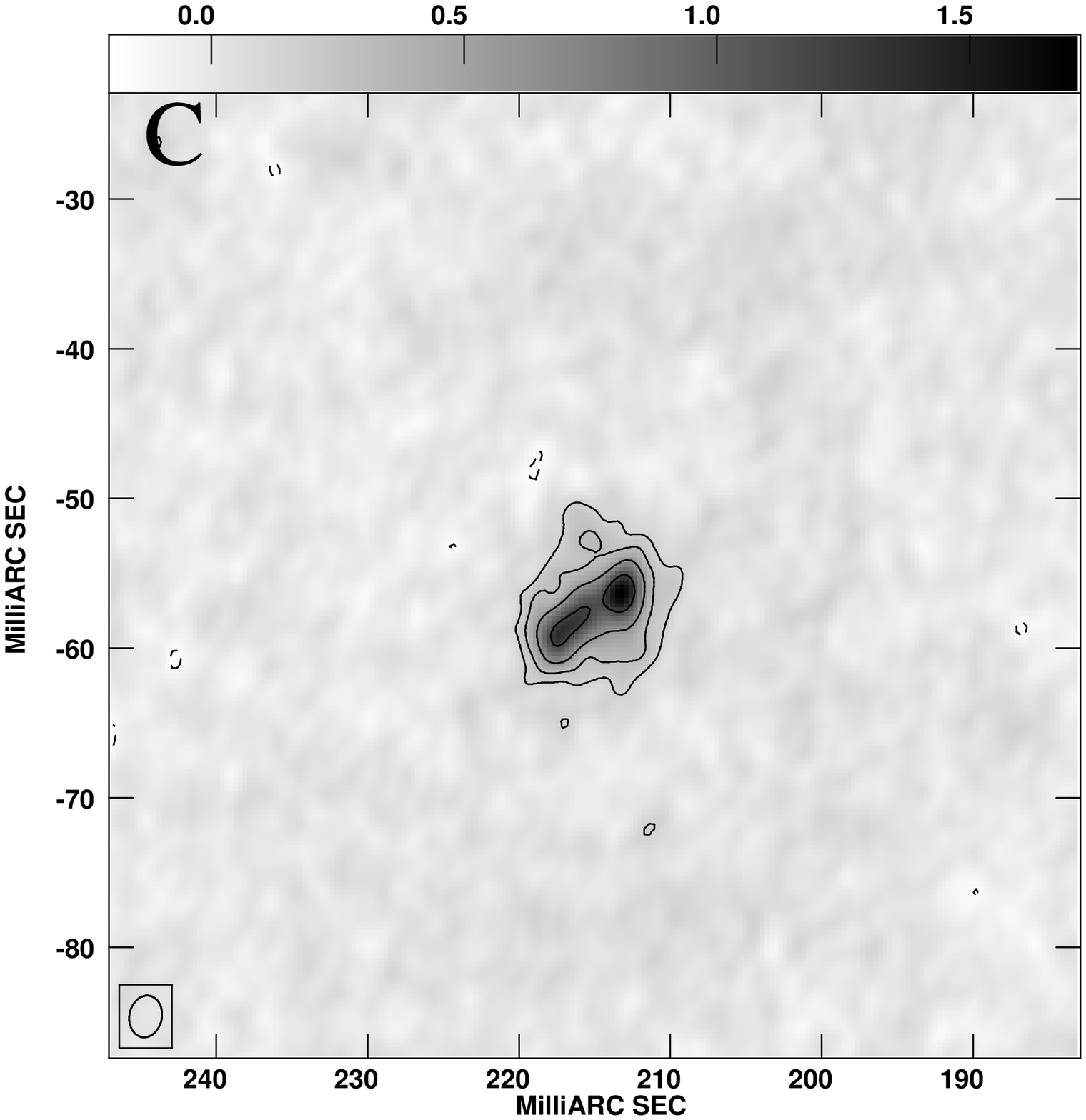}}
\resizebox{0.3\hsize}{!}{\includegraphics{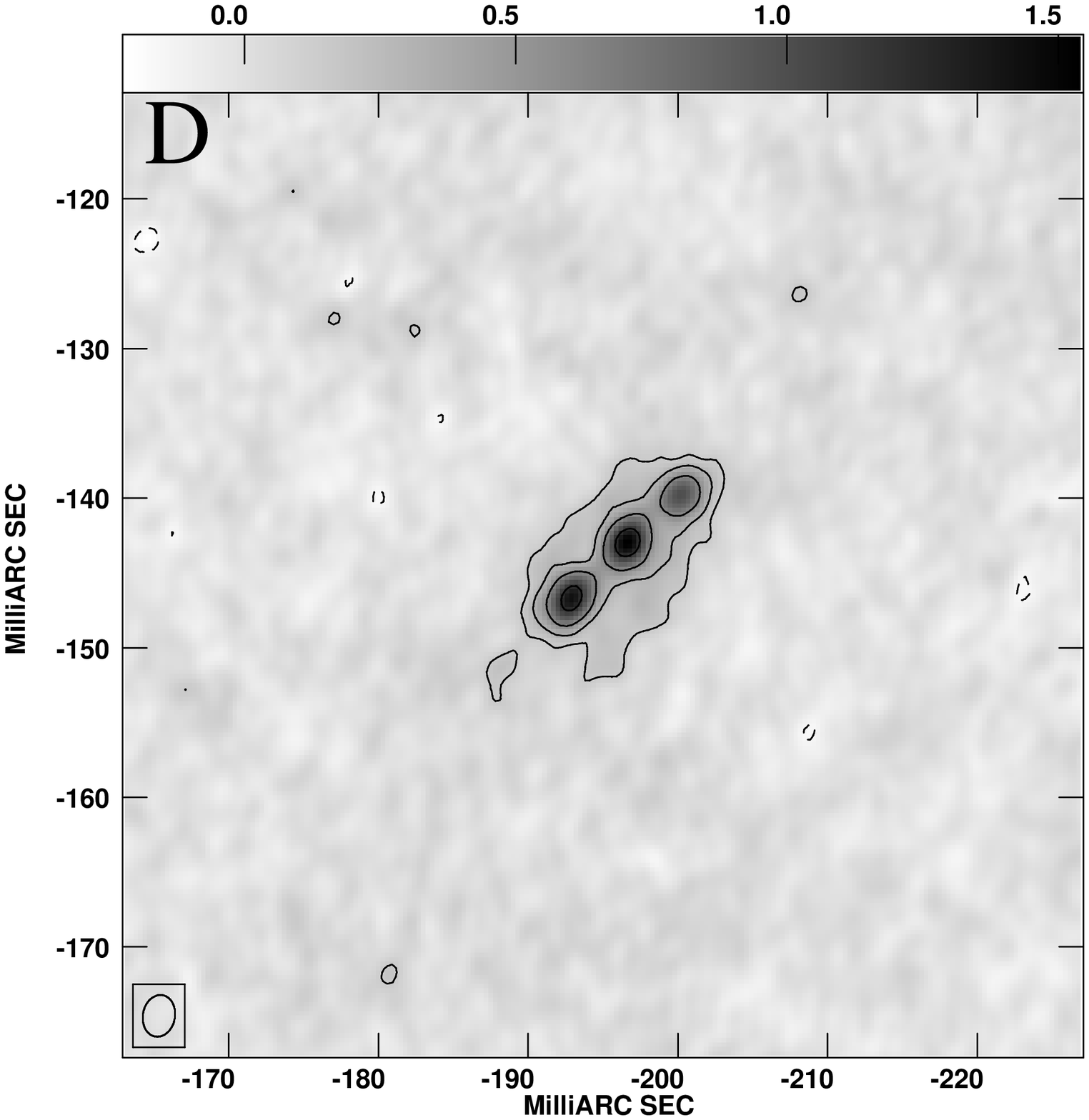}}
\parbox[b]{0.37\hsize}{
\caption{\small Very-Large Baseline Array (VLBA) radio images of
B0128+437 at 5\,GHz. Whereas images A, C and D clearly show similar
images structure, consisting of three distinctive knots, in image B
these knots are nearly gone. Since lensing conserves surface
brightness, such dramatic changes are most easily explained by strong
scattering due to the ionized lens ISM. With SKA one could
similarly probe the ionized ISM of high-z galaxies.\label{fig:0128}}\smallskip\medskip}}}
%\end{center}
\end{figure*}

\section{Other Lensing Studies with SKA}

Besides an extensive quantitative study of the internal structure and
evolution of galaxies and clusters to $z\sim 1$, a large sample of
new radio-bright lens systems allows more ``exotic'' studies:

\begin{itemize}

\item {\bf Time--Delays:} Intrinsic brightness and structural changes
in the lens source will induce correlated variability in the observed
lensed images. The different traverse times of photons for these
images introduce time-delays that can be measured and used to
determine the Hubble constant if the lens mass model is
known. However, once SKA comes on line, this and the other
cosmological parameters will have been determined with great accuracy
already. Time-delays can then be used -- given a fixed cosmological
model -- as a powerful direct measure of the density profile of
individual galaxies. A combination of potentially hundreds or
thousands of time-delays allow a precise (few percent) determination
of the average radial density profile in their inner $\sim$15\,kpc. To
select lensed sources that are also variable, the RASKAL Survey has to
be designed to make {\sl multiple} passes over the same area. After an
initial selection of the brightest and most variable lensed sources,
several tens of sources can simultaneously be monitored using SKA's
multi-beam capability (if this design is implemented)\footnote{In
1\,sec, a S/N$\sim$100 is reached on a 1\,mJy source. With one beam,
one could monitor {\sl all} 1-mJy sources in the sky in 6 hrs (i.e.\
600 times less than RASKAL itself), or in 30 min with ten 1-sq. degree
beams.}.

\medskip

\item {\bf Micro \& Milli--lensing:} Small scale structure in the lens
potential -- e.g. due to luminous mass (stars, globular clusters,
etc.) and, possibly, CDM substructure -- can cause differential
magnification of the source, on milli- to micro-arcsec scales, and
deviations of the lensed image properties (e.g.\ positions and
flux-ratios) from those naively expected from a smooth lens
potential. If the structure of the source or the potential change, or
their relative alignment on the sky (due to deviations of the lens
and/or source velocities wrt the Hubble flow), apparent structural
source variability that is uncorrelated between images, can occur.
Through this, one can study both the small-scale properties of the
lens potential (e.g.\ their stellar or CDM substructure content) and
the radio-source structure on scales unattainable through other
methods. With a statistical study of the {\sl structure functions} of
thousands of the bright and compact lensed AGNs, one might e.g. be
able to disentangle the power-spectra of the source structure from
that of the lens potential. If the latter is dominated by stars, such
power-spectra can be used to constrain the stellar mass function in
cosmologically-distant galaxies.

\medskip

\item {\bf The Lens--Galaxy ISM:} Radio-wave scattering due to the
ionized ISM in the lens galaxy can affect the lensed images (e.g.\
scatter-broadening and scintillation; Figure 3).
The time-scales and frequency dependence of the source variability (both
in structure and flux) will be different for milli/microlensing
(lensing is achromatic)\footnote{Whereas intrinsic variability can be
removed because it is correlated between the lensed images, scattering
and milli- and microlensing are uncorrelated.}. By studying the
frequency dependent behavior of the lensed images -- in particular if
the source is scatter-broadened on scales of $\sim$1\,mas -- one might
be able to constrain the shape and normalization of the power-spectrum
of the ionized ISM in the lens galaxies. Hence, scattering and milli-
and microlensing are very similar, with the main difference being that
in the former case the ``scattering screen'' is frequency
dependent. The mathematical toolbox developed to study
scintillation/scattering can thus be applied to milli- and
microlensing as well. Since the ionized ISM is related to star
formation, such studies can provide a gauge of the amount of star
formation in highly-obscured star-burst galaxies. In addition,
polarized background sources that are lensed into extended arcs or
even rings can be used, through Faraday rotation, to map $\int n_{\rm
e} B_{\parallel} {\rm d}r$ in the lens itself. In the case of
clusters, both X-ray and S-Z observations can constrain the electron
density $n_{\rm e}$. Hence their large-scale coherent magnetic field,
$B_{\parallel}$, can be quantified and mapped.

\medskip

\item {\bf High--Redshift Sources:} The use of strong lensing as a
``natural telescope'', magnifying faint high-redshift source is
becoming more important. Source magnifications of $\mu\sim 10$ for
example ``upgrade'' 8--10m class telescopes to a 30m telescope!  The
magnification of strongly-lensed images that merge near a critical
curve is $\mu \sim {\theta_{\rm Einst}/\Delta \theta_{12}}$, where
$\Delta \theta_{12}$ is the angular distance between the two merging
images and $\theta_{\rm Einst}$ the Einstein radius of the
lens. Whereas ground-based optical images limit $\Delta \theta_{12}$
to $\sim$1$''$ and $\mu\sim 30$ for typical clusters with $\theta_{\rm
Einst}\sim30''$, the high resolution of SKA with a limit of $\Delta
\theta_{12}\sim0.05''$ allows radio sources with $\mu\sim 600$ to be
found and studied, upgrading SKA to an effective Thousand Square
Kilometer Array! Not only does this allow extremely faint sources at
high redshift to be found, through the high magnification (mostly
linear) details in the source can be studied that otherwise can only
be resolved with $10^6$~km baselines. We expect this technique to
provide the most detailed studies of faint high-$z$ radio sources,
possibly the progenitors of AGNs and other faint radio sources at or
beyond the epoch of reionization. Per cluster the lensing
cross-section, however, is $\sigma(>\mu) \sim \pi \theta^2_{\rm
Einst}/\mu^2$. If the source count increases as $N(>S_{\rm obs},
\mu)\sim 3\times 10^5 (S_{\mu{\rm Jy}}/\mu)^{-1}$ per square degree,
the number of highly-magnified lensed sources per cluster is $N_l\sim
0.1 (\theta_{\rm Einst}/1'')^2 (\mu S_{\mu{\rm Jy}})^{-1}$.  We might
expect hundreds of highly magnified source (with $\mu$$>$100) from the
thousands of clusters observed in a RASKAL Survey.

\medskip

\item {\bf Rare Lensing Events:} Besides these exciting possibilities,
the large number of lens systems will naturally produce rare and odd
cases of lensing: (1) Lensing by higher-order catastrophes (e.g.\
hyperbolic umbilics, swallow-tails, etc.), for example, provide large
source magnifications and interesting constraints on the lens
potential. (2) The radio-afterglows of GRBs can be multiply imaged by
foreground galaxies. This not only provides accurate image time
delays, but also an opportunity to re-direct {\sl optical, X--ray and
other} telescopes to ``the place of action'' {\sl before} the burst in the
other lensed images occur (because of the time-delays). (3)
Microlensing of GRB bursts, while multiply-imaged, will further allow
a probe into their $\mu$as-scale structure during the initial
phases. (4) The large number of systems also allows rare cases of
extremely small and extremely large image separations to be found by
dwarf galaxies and massive clusters, respectively. (5) Although shown
to be rare already, collapsed {\sl dark structures} that have no
luminous component, can also be found through strong lensing in an
unambiguous way if they really exist in significant numbers.

\end{itemize}

\section{The Synergy with Optical Telescopes}

To fully exploit the unique capabilities of SKA to discover and study
strong gravitational lenses, it is essential to follow-up a
significant fraction of these systems at optical wavelengths. The
reasons are many-fold: (i) Identification of the lens type (e.g.\
early- versus late-type galaxies, or other), something that is not always
obvious from only the radio observations. (ii) Characterization of the
stellar mass distribution, its colors and luminosity. This is
particularly important if a serious comparison between luminous and
dark-matter properties is planned. (iii) Determination of photo-
and/or spectroscopic redshift of the lens and source. (iv)
Spectroscopic follow-up to study the kinematics of the lens and its
chemical constituents (incl. dust-extinction of the lensed images). (v)
Study the highly magnified and high-$z$ source, using the lens as a
natural magnifying glass. These are only a few reasons, but it is
clear that optical follow-up is needed to fully exploit these lens
systems in the context of cosmology, galaxy formation and
evolution.

The study of arc-second strong lensed systems has greatly benefited from the
high resolution offered by the {\sl Hubble Space Telescope}. With
$>10^5$ lensed systems, however, individual follow-up will not be feasible and
deep all or large sky surveys in the optical are the only
option. Several instruments are planned that strive toward this,
among which are the {\sl Large-aperture Synoptic Survey Telescope}
(LSST) and the {\sl Panoramic Survey Telescope \& Rapid Response
System} (Pan--STARRS). These are ground-based instruments that intend
to monitor the entire visible sky, although limited in resolution by
atmospheric seeing ($>$0.5$''$). The space-based {\sl Supernova
Acceleration Probe} (SNAP) will image smaller areas of the sky (up to
$\sim$300 sq. degrees) with higher resolution (0.1$''$). Powerful
deconvolution techniques, however, can be applied to low-resolution,
but high S/N, ground-based images to extract the necessary optical
information.

With SKA, we expect to find $>10^{5-6}$ new lensed systems, have $0.5-1.0''$
ground-based optical images of most of these and $0.1''$ resolution
space-based optical images of $\sim 10^{3-4}$ systems (several
percent). Spectroscopic follow-up can probably be done of a similar
number of systems with large field-of-view multi-fiber spectrograph's
on large-aperture optical telescopes.  In particular for kinematic
studies this requires at least 8--10m class telescopes, most
preferably larger, such as the next-generation of {\sl Extremely Large
Telescopes} (ELTs; i.e. $>20$\,m). Even though essential, the
bottleneck of a serious strong-lensing study of the majority of new
lens systems that SKA will discover is the optical (or other
wavelength) follow-up.  This is not only true for lensing studies, but
applies to most high--$z$ studies where objects are typically $<1''$
in size.  A serious study is therefore needed how to match optical
surveys to the capabilities of SKA, in depth, resolution and
sky-coverage, fully exploiting their complementary capabilities.

\section{Prospects}

The state of lensing is good! And it will drastically improve once SKA
comes on line.  However, a new telescope is not only built to redo
``old stuff'' (i.e.\ old once SKA is build). With a sample of
$10^{5-6}$ strong-lens systems one can explore a much wider range of
lens-galaxy or cluster properties (e.g.\ in mass, luminosity,
redshift, etc.). It is hard to break up current samples of a few dozen
useful lensed systems in more than a few bins.

The {\sl Square Kilometer Array} will be hard to beat in terms of
finding lensed systems, since it has superior resolution over all
planned optical telescopes -- which also aim to find lensed systems --
and the lenses themselves are often faint at radio wavelengths,
limiting confusion with the lensed source.  However, there is no
guarantee that lensed radio sources are also optically bright, which
is something that needs to be kept in mind. Even if they are, one can
only find out by optical follow-up.  This requires a similar all-sky
survey in the optical, as planned with LSST and
Pan-STARRS. Determining most lens and source redshifts (if not
possibly from HI emission), requires spectroscopic follow-up with
large field-of-view telescope of at least 8--10m diameter. Because of
their more limited sample size, cluster-studies will be less affected
by limits on optical follow-up. We expect these issues to be addressed
and solved, since they are crucial in many of the studies with SKA.

Once these issues are addressed, SKA can revolutionize the study of
the internal structure and evolution of galaxies and clusters through
extremely detailed lensing studies in combination with complementary
optical (and possibly X-ray) data of large samples. There is currently
no obvious ``competitive'' technique that can do this in an
unambiguous way (e.g.\ HI kinematic studies are limited mostly to
gas-rich late-type galaxies).  In addition SKA will open up new fields
of study at cosmological distances, such as measuring the ionized ISM,
the mass function of stars and possibly CDM substructure in high-$z$
galaxies.  The resolution and sensitivity of SKA will also allow
extremely magnified high-$z$ sources to be found when lensed by
clusters (and galaxies alike), providing unique insight into the
structure of the faintest and highest redshift population of radio
sources that even with SKA are beyond reach when not magnified.
Besides the obvious, a large sample will undoubtedly show up more
exotic types of lensing (e.g.\ higher-order catastrophes, lensing of
GRBs, etc.) which are just too rare to find with the present limited
capabilities and samples.

{\sl We conclude to say that through strong gravitational lensing, SKA
can have a major impact upon numerous fields in cosmology and
astrophysics and undoubtedly open up many new and unexpected avenues
of research.}

\end{document}